\documentclass[9pt,twocolumn,superscriptaddress,floatfix,pra]{revtex4-1}

\usepackage{amsthm}
\usepackage{amsmath}
\usepackage{bbm, dsfont}
\usepackage{graphicx}

\newcommand{\be}{\begin{equation}}
\newcommand{\ee}{\end{equation}}

\newcommand{\del}[1]{}
\newcommand{\appen}{Appendix}

\begin{document}
\title{Optically probing the detection mechanism in a molybdenum silicide superconducting nanowire single-photon detector\vspace{.5cm}} 
\author{Misael Caloz}\email{misael.caloz@unige.ch}
\affiliation{Group of Applied Physics, University of Geneva, Chemin de Pinchat 22, CH-1211 Geneva 4 Switzerland}
\author{Boris Korzh}
\affiliation{Group of Applied Physics, University of Geneva, Chemin de Pinchat 22, CH-1211 Geneva 4 Switzerland}
\author{Nuala Timoney}
\affiliation{Group of Applied Physics, University of Geneva, Chemin de Pinchat 22, CH-1211 Geneva 4 Switzerland}
\author{Markus Weiss}
\affiliation{Department of Physics, University of Basel, Klingelbergstrasse 82, CH-4056 Basel Switzerland}
\author{Stefano Gariglio}
\affiliation{Department of Quantum Matter Physics, University of Geneva, 24, Quai Ernest-Ansermet, CH-1211 Geneva 4 Switzerland}
\author{Richard J.~Warburton}
\affiliation{Department of Physics, University of Basel, Klingelbergstrasse 82, CH-4056 Basel Switzerland}
\author{Christian Sch\"onenberger}
\affiliation{Department of Physics, University of Basel, Klingelbergstrasse 82, CH-4056 Basel Switzerland}
\author{Jelmer Renema}
\affiliation{Clarendon Laboratory, University of Oxford, Parks Road, Oxford OX1 3PU United Kingdom}
\author{Hugo Zbinden}
\author{F\'elix Bussi\`eres} \email{felix.bussieres@unige.ch}
\affiliation{Group of Applied Physics, University of Geneva, Chemin de Pinchat 22, CH-1211 Geneva 4 Switzerland}

\begin{abstract}
\vspace{0.5cm}

We experimentally investigate the detection mechanism in a meandered molybdenum silicide (MoSi) superconducting nanowire single-photon detector by  characterising the detection probability as a function of bias current in the wavelength range of 750 to 2050~nm. Contrary to some previous observations on niobium nitride (NbN) or tungsten silicide (WSi) detectors, we find that the energy-current relation is nonlinear in this range. Furthermore, thanks to the presence of a saturated detection efficiency over the whole range of wavelengths, we precisely quantify the shape of the curves. This allows a detailed study of their features, which are indicative of both Fano fluctuations and position-dependent effects.

\vspace{1.0cm}
\end{abstract}
\maketitle
Superconducting nanowire single-photon detectors (SNSPDs) are a key technology for optical quantum information processing~\cite{Goltsman2001a, Hadfield2009}. An SNSPD consists of a thin wire of superconducting material biased close to its critical current, which becomes resistive after the absorption of a single photon, leading to a detection through an amplified voltage pulse. Their low dark count rate, fast response time, small jitter, and high efficiency favours their use in various demanding quantum optics applications such as quantum key distribution~\cite{Takesue2007}, quantum networking~\cite{Bussieres2014}, device-independent quantum information processing~\cite{Shalm2015a} and deep-space optical communication~\cite{Shaw14}. Notably, SNSPDs can be integrated in photonic circuits~\cite{Sprengers2011,Rath2015}. 

One recent important advance in the SNSPD field has been the introduction of amorphous superconductors such as tungsten silicide (WSi)~\cite{marsili2013}, molybdenum silicide (MoSi)~\cite{korneeva2014,Verma2015} and molybdenum germanium (MoGe)~\cite{Verma2014b}. SNSPDs based on these materials currently have the highest reported detection efficiencies (93\% for WSi~\cite{marsili2013}), as well as a higher fabrication yield~\cite{Allman2015a} than devices made of polycrystalline materials such as niobium nitride (NbN)~\cite{Goltsman2001a}, niobium titanium nitride (NbTiN)~\cite{Dorenbos2008} and tantalum nitride (TaN)~\cite{Engel2012TaN}. 

One striking difference with polycrystalline materials is that amorphous SNSPDs have a detection efficiency that saturates at bias currents well below the critical current~\cite{Baek2011a}. Despite extensive studies, the question remains if these differences are due to a fundamentally different detection mechanism. Moreover, understanding the nature of the detection mechanism may ultimately lead to novel SNSPD structures with better performances or SNSPD-inspired devices targeting a broader range of applications.

One of the main techniques for investigating the detection mechanism is measurements of the energy-current relation, i.e.\@ the amount of photon energy required to produce a detection event at constant detection probability. For NbN, the energy-current relation was found to be linear~\cite{Renema2014} over a large range of energies using quantum detector tomography~\cite{Akhlaghi2011} (QDT), which is evidence for the role of a diffuse cloud of quasiparticles in the detection process. Moreover, position-dependent measurements~\cite{Renema2015} and external magnetic field-based study~\cite{Vodolazov2015} highlight the role of vortices in the detection mechanism. In WSi, a linear relation was found over a large range of energies, but with a slight deviation from a linear behaviour at low energies. Other results are, however, contradictory: in separate experiments, indications of a nonlinear energy-current relation were found for NbN and WSi SNSPDs~\cite{Vodolazov2015}. In contrast, no extensive studies have been carried out on amorphous MoSi devices. 

In this work, we experimentally investigate the detection mechanism in MoSi SNSPDs. We illuminate a 170~nm wide MoSi SNSPD with wavelengths ranging from 750 to 2050~nm. By recording the photon count rate as a function of the bias current and the incident photon energy, we are able to characterise fully the device response. We find that the energy-current relation is nonlinear throughout this wavelength range. Furthermore, we investigate the shape of the count rate curves at different photon energies. We interpret these results as a potential combination of Fano fluctuations and position-dependent effects in the device.

\begin{figure}[t!] 
	\includegraphics[width=85mm]{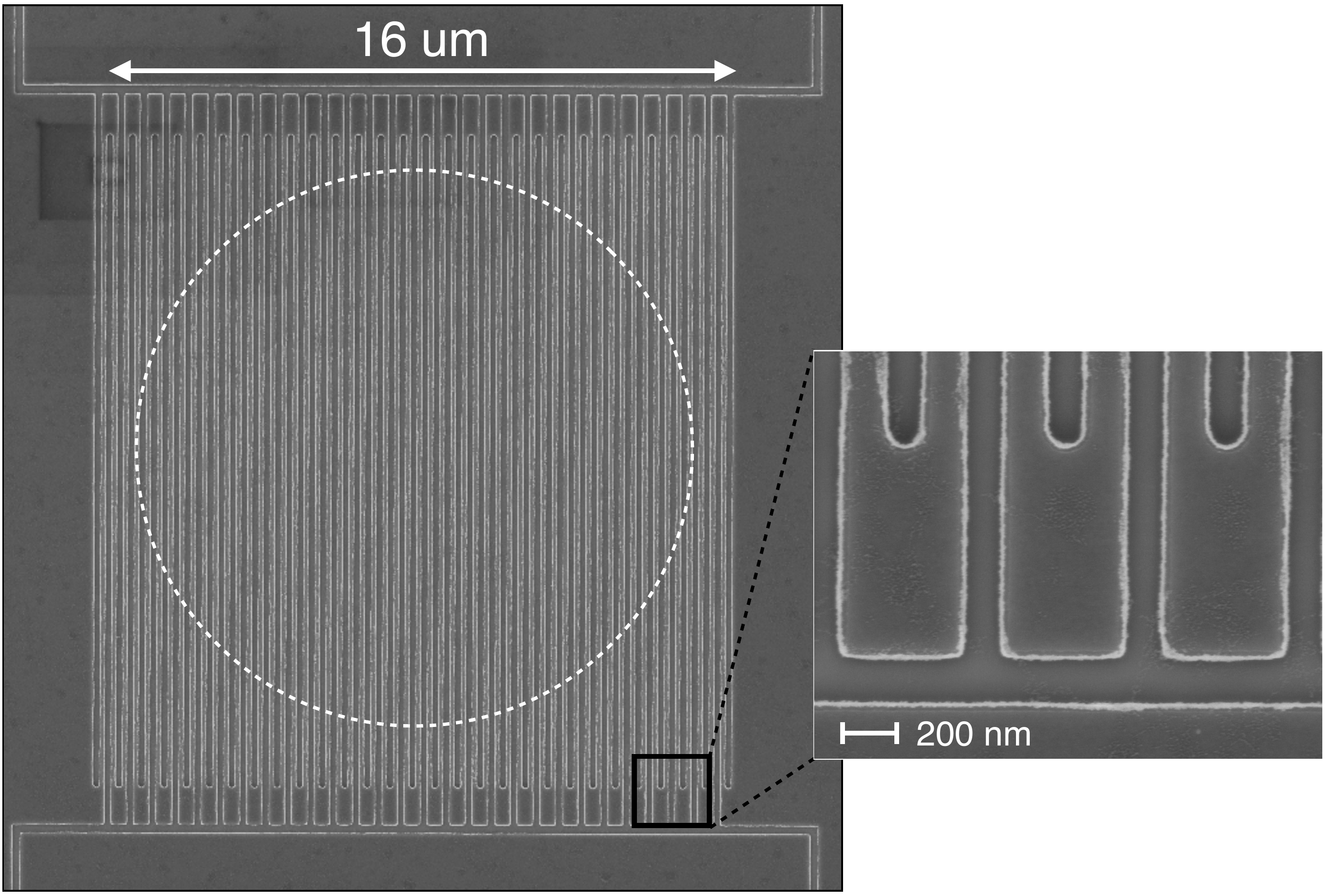}
	\caption{Scanning electron microscope image of the MoSi device. The dashed circle shows where the photons are absorbed, corresponding to the limit of the gaussian mode spread from the optical fiber. The inset on the right shows a magnification of the meander turns.}\label{fig:meander}
\end{figure}

The device is fabricated out of a 5~nm thick film of amorphous Mo$_{0.8}$Si$_{0.2}$, with a $T_c = 5$~K, which is deposited by co-sputtering Mo and Si targets. The film is patterned into a meandered wire with a width of 170~nm and a pitch of 160~nm (see Fig.~\ref{fig:meander}) and a total surface area of 16$\times$16~$\mathrm{\mu}$ m$^2$ by a combination of e-beam lithography and ion beam etching. A self-aligning technique is used to ensure optimal coupling to the optical fibre~\cite{Miller2011}. The device has been selected out of tens of other detectors by looking at the highest critical current and widest plateau region. The detection efficiency at 1550~nm is 20\%.

The detector is mounted in a sorption cryostat reaching 0.75 K. The detector is biased with a current source and its critical current is 14.7~$\mathrm{\mu}$A. The voltage pulses from detection events are amplified by a custom low-noise amplifier cooled to 40~K and by a secondary amplifier at room temperature.

The detector is illuminated with unpolarized photons coming from a halogen lamp sent through a grating monochromator. This provides a continuous spectrum from 750 to 2050~nm. We carefully calibrated the monochromator using laser lines at 632.8, 980.1, 1064.0, 1310.2 and 1550.8~nm. By using the second order of some of these wavelengths, we obtain 9 calibration points, extending up to 2128.0 nm with a 4~nm uncertainty. Appropriate low pass filters were inserted to avoid crosstalk from higher diffraction orders. 

\begin{figure}[t!] 
	\includegraphics[width=85mm]{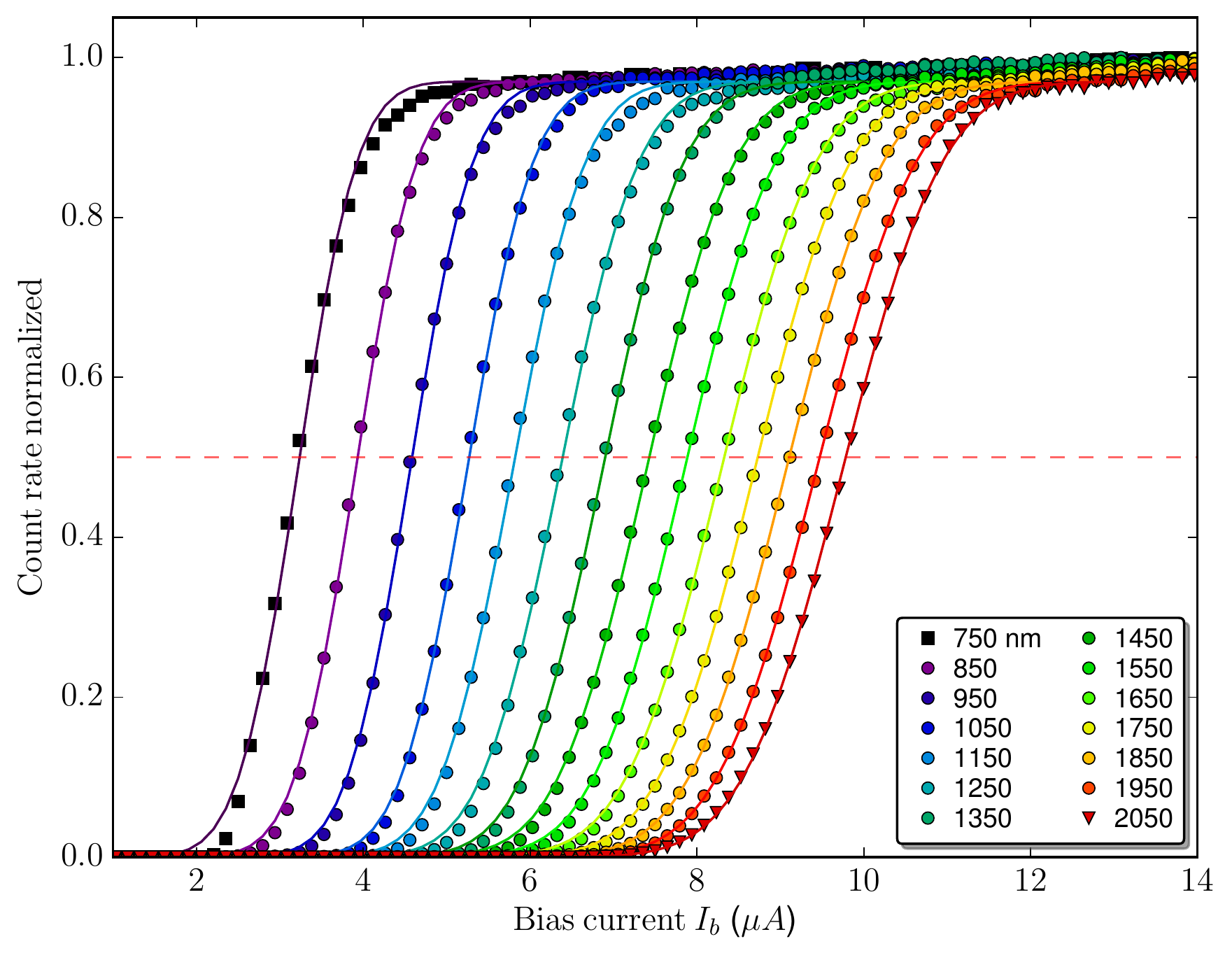}
	\caption{Normalized photon count rate (detection rate subtracted minus the DCR, normalized by the maximum count rate) as a function of the bias current $I_b$ at 0.75~K. Each color represents one measurement run with a specific incident photon wavelength. Each solid line traces the error function fit for the respective data curve. The dashed red line indicates the fraction $\eta$ of the saturated detection efficiency $\eta= 50\%$. The leftmost and rightmost curves correspond to 750 and 2050~nm, respectively. The critical current is 14.7~$\mathrm{\mu}$A}\label{fig:energy-scan}
\end{figure}

We measured the photon count rate (PCR) as a function of bias current and photon energy, integrating for 10~seconds at each point; see Fig.~\ref{fig:energy-scan}. We measured the system dark count rate (DCR) and subtracted it from our measurements. In order to compare various wavelengths, we normalise our data to a count rate value situated just below the critical current, i.e.\@ in the plateau region where the efficiency is saturated. 

To reconstruct faithfully the curves one must pay attention to the pulse discrimination electronics. 
Indeed, a problem can arise when the detector operating at very low bias currents, i.e.~at currents for which the amplitude of the detection pulses are marginally higher than the amplitude of the noise of the amplifying chain and of the discriminator level. The consequence is that the shape of the PCR curve can be affected. We avoid this problem by operating only at those currents and discriminator levels where the shape of the curves are independent of the discriminator level. See the \appen{} for details.

\begin{figure}[t!] 
	\includegraphics[width=85mm]{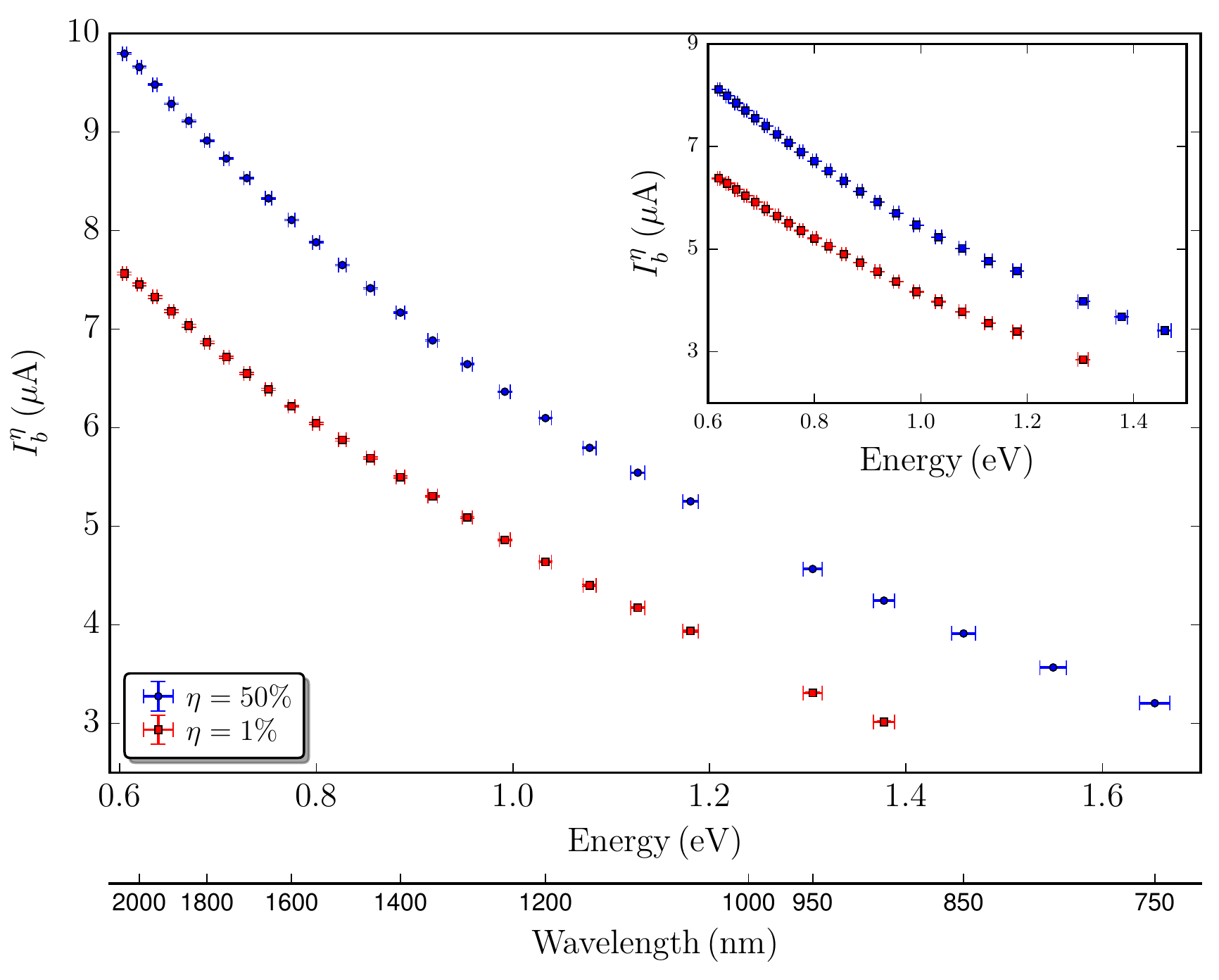}
	\caption{Energy-current relation for two different normalised detection probabilities. The threshold current $I_{b}^{1\%}$ (red squares) and $I_{b}^{50\%}$ (blue points) are plotted as a function of the photon energy and corresponding wavelength, at 0.8 K. Inset: Energy-current relation at 1.5 K.}\label{fig:energy-current}
\end{figure}

Fig.~\ref{fig:energy-current} shows the energy-current relation for our MoSi detector. For each wavelength we plot the amount of bias current $I_b^{\eta}$ required to achieve a certain fraction $\eta$ of the saturated detection efficiency (which we normalised to one in Fig.~\ref{fig:energy-scan}). Our setup allows us to measure from 0.6~eV to more than 1.6~eV in the single-photon absorption regime. We plot this relation for $\eta = 50\%$ and $\eta = 1\%$, at 0.8~K and 1.5~K. We find that the relation between bias current and photon energy is nonlinear throughout this entire measurement range, and for both temperatures. 

The long plateau and the broad response of our detector allows us to carefully characterise the full shape of the normalised PCR curves, and compare them with models in the literature. The curves have a transition region where the detection efficiency increases, followed by a plateau region. One theory attributes the shape of the transition region to Fano fluctuations, which are the result of the statistical nature of the quasiparticle creation process~\cite{Fano1947, Kozorezov2008}. Since only a finite fraction of the incoming photon energy ends up in the quasiparticle bath, the number of quasiparticles generated by a photon of energy $E$ fluctuates as $\Delta N = \sqrt{FE/\epsilon}$, where $F$ is the Fano factor and $\epsilon$ is the energy of a single quasiparticle. These fluctuations have recently been analysed in the context of a model of quasiparticle recombination~\cite{Kozorezov2015}. In this model, the transition region occurs because for some currents, the photon only occasionally produces enough quasiparticles to trigger a detection. This results in a predicted sigmoidal shape (error function) for the PCR curve with a width that is set by the microscopic details of the downconversion process~\cite{Kozorezov_SPW}.

\begin{figure}[t!] 
	\includegraphics[width=85mm]{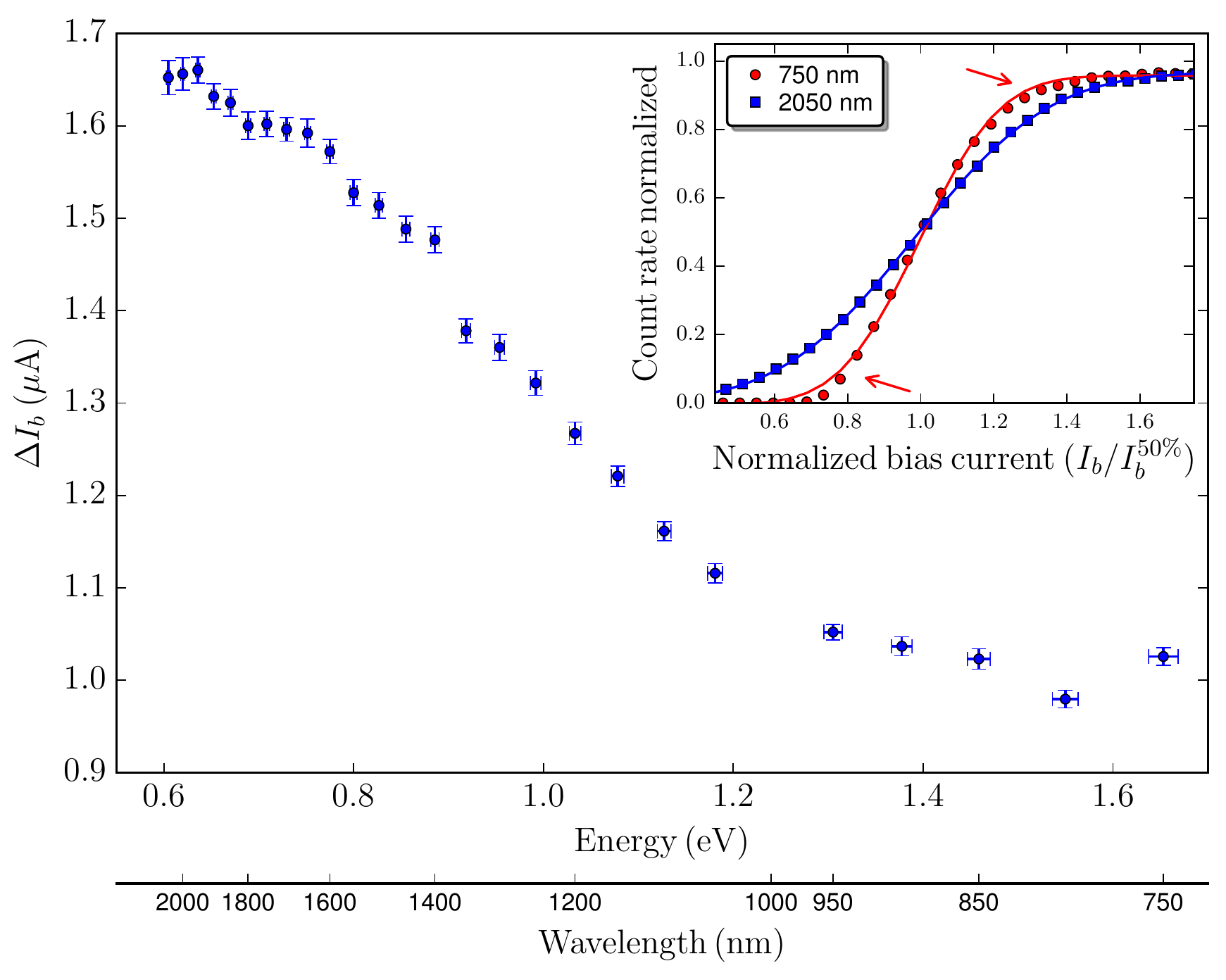}
	\caption{Transition width defined as $\Delta I_b = I_{b}^{80\%} - I_{b}^{20\%}$ obtained from Fig.~\ref{fig:energy-scan} as a function of the incident photon energy. Inset: photon count rate curve for 750~nm and 2050~nm as a function of the normalised bias current. The solid lines represent the error function fit, see text. The red arrows indicate the two inflection points where the data is not well described by the error function.}\label{fig:energy-sigma}
\end{figure}

To check whether the Fano fluctuation theory agrees with our measurements, in Fig.~\ref{fig:energy-scan} we fit the experimental data with an error function $R(I_b) = \mathrm{erf}\left[{(I_b - I^{50\%}_b)/{\sigma \sqrt{2}}}\right]$, where $\sigma$ quantifies the width of the transition. As can be seen, at low photon energies the fit agrees very well with the data. However, at high energies, the shape of the curves starts to deviate from the $R(I_b)$ fits. The inset in Fig.~\ref{fig:energy-sigma} shows the highest and lowest energy scans, which are overlapped to facilitate comparison. This discrepancy is statistically significant: the difference in the reduced $\chi^2$, which quantifies the quality of the fit, is over two orders of magnitude between the lowest and highest photon energies. See \appen{} for details.

From Fig.~\ref{fig:energy-scan}, it is clear that the transition becomes narrower as the photon energy is increased. Figure~\ref{fig:energy-sigma} shows the width of the transition as a function of photon energy, defined as $\Delta I_b = I_{b}^{80\%} - I_{b}^{20\%}$. While this effect is observed in previous studies~\cite{Baek2011a,Engel2012TaN}, we believe we present here its first quantitative description. The interpretation of this effect is still an open problem.

The error function fit which we observe for low photon energies is precisely what is predicted by the Fano model. However, deviations from this shape at high energies suggests that this may not be the whole story. A possible explanation could come in the form of an additional model which predicts position dependent effects in the nanowire. In this model, different parts of the cross section of the superconducting nanowire become photodetecting at different bias currents, due to an intrinsic position dependence in the fundamental detection mechanism~\cite{Renema2015, Vodolazov2014}. In such a model, different points in the cross-section of the wire have different energy-current relations. Consequently, this gives rise to additional broadning of the transition (in addition to the Fano fluctuations), where the width of the transition is given by $\Delta I_b = I_{\text{min}}(E)-I_{\text{max}}(E)$, where $I_{\text{min}}$ and $I_{\text{max}}$ are the threshold currents at the most efficient point (edge) and the least efficient point (middle) along the cross-section of the wire, respectively. For such a model, one expects the width of the transition to increase with higher photon energies~\cite{Renema2015b, Renema2015}, which could explain why the error function fit is not as good at higher photon energies. Moreover, due to the sharpening of the error-function transition (Fig.~\ref{fig:energy-sigma}) at higher photon energies, one would expect any additional effects to be more visible, even if the position dependence effect is weakly dependent on photon energy.

We note that due to the transition width energy dependence shown in Fig.~\ref{fig:energy-sigma}, the probability threshold of the energy-current relation $\eta= 50\%$ is a good choice. Indeed, the relation for $\eta = 1\%$ is affected by the transition width dependence and appears closer to linear. We observe this effect by comparing both curves in Fig.~\ref{fig:energy-current}, but it does not change qualitatively the non-linear behaviour.

The nonlinear relation in MoSi is surprising in the light of previous experiments. For 220~nm-wide NbN SNSPDs made from nanobridges, and also with nanodetectors and meanders, the energy-current relation was found to be linear in the range of 0.75 to 8.26~eV using quantum detector tomography (QDT)~\cite{Renema2014}. A result consistent with this was found for TaN detectors~\cite{engel2013numerical}, and for a series of NbN meanders of varying widths~\cite{Lusche2014}. Nevertheless, a nonlinear behaviour for NbN meanders probed with a filtered black body light source was later observed in the 0.5 to 2.75~eV range~\cite{Vodolazov2015} by using the two probability thresholds $\eta = 50\%$ and $90\%$ of the normalised PCR. For the amorphous materials, the evidence is scarcer: a previous study with WSi meanders found a linear relation at low energies and a single point deviating from this trend at 1.8~eV~\cite{Baek2011a}. Recently, measurements on 220~nm-wide WSi SNDPDs nanobridges~\cite{Gaudio2016} using QDT have shown a linear behaviour from 0.85 to 2.5~eV, but with a slight deviation from the linear behaviour between 0.75 and 0.85~eV. Reviewing this seemingly contradictory evidence, no obvious distinction between the two groups of results presents itself: neither wire width, nor device geometry, nor measurement method, nor the crystallinity of the material. While our results add additional data, the question of the detection mechanism remains an open problem.


In conclusion, we investigated the detection mechanism in MoSi superconductor nanowires single-photon detectors by measuring the PCR as a function of photon energy and bias current. We found a nonlinear energy-current relation, in contrast to some observations on other materials such as NbN and WSi. Moreover, we study the full shape of the detection probability curve and found indications for the role of both Fano fluctuations and position-dependent effects.


The authors would like to acknowledge Claudio Barreiro and Daniel Sacker for technical assistance, A. Kozorezov, D.Y. Vodolazov, V. Verma and F. Marsilli for scientific discussions, and the Swiss NCCR QSIT (National Center of Competence in Research - Quantum Science and Technology) for financial support. This work was partly supported by the COST (European Cooperation in Science and Technology) Action MP1403 - Nanoscale Quantum Optics.


\bibliography{Geneva_MoSi_SNSPDs_arxiv.bbl}

\begin{thebibliography}{30}%
\makeatletter
\providecommand \@ifxundefined [1]{%
 \@ifx{#1\undefined}
}%
\providecommand \@ifnum [1]{%
 \ifnum #1\expandafter \@firstoftwo
 \else \expandafter \@secondoftwo
 \fi
}%
\providecommand \@ifx [1]{%
 \ifx #1\expandafter \@firstoftwo
 \else \expandafter \@secondoftwo
 \fi
}%
\providecommand \natexlab [1]{#1}%
\providecommand \enquote  [1]{``#1''}%
\providecommand \bibnamefont  [1]{#1}%
\providecommand \bibfnamefont [1]{#1}%
\providecommand \citenamefont [1]{#1}%
\providecommand \href@noop [0]{\@secondoftwo}%
\providecommand \href [0]{\begingroup \@sanitize@url \@href}%
\providecommand \@href[1]{\@@startlink{#1}\@@href}%
\providecommand \@@href[1]{\endgroup#1\@@endlink}%
\providecommand \@sanitize@url [0]{\catcode `\\12\catcode `\$12\catcode
  `\&12\catcode `\#12\catcode `\^12\catcode `\_12\catcode `\%12\relax}%
\providecommand \@@startlink[1]{}%
\providecommand \@@endlink[0]{}%
\providecommand \url  [0]{\begingroup\@sanitize@url \@url }%
\providecommand \@url [1]{\endgroup\@href {#1}{\urlprefix }}%
\providecommand \urlprefix  [0]{URL }%
\providecommand \Eprint [0]{\href }%
\providecommand \doibase [0]{http://dx.doi.org/}%
\providecommand \selectlanguage [0]{\@gobble}%
\providecommand \bibinfo  [0]{\@secondoftwo}%
\providecommand \bibfield  [0]{\@secondoftwo}%
\providecommand \translation [1]{[#1]}%
\providecommand \BibitemOpen [0]{}%
\providecommand \bibitemStop [0]{}%
\providecommand \bibitemNoStop [0]{.\EOS\space}%
\providecommand \EOS [0]{\spacefactor3000\relax}%
\providecommand \BibitemShut  [1]{\csname bibitem#1\endcsname}%
\let\auto@bib@innerbib\@empty
\bibitem [{\citenamefont {Gol'tsman}\ \emph {et~al.}(2001)\citenamefont
  {Gol'tsman}, \citenamefont {Okunev}, \citenamefont {Chulkova}, \citenamefont
  {Lipatov}, \citenamefont {Semenov}, \citenamefont {Smirnov}, \citenamefont
  {Voronov}, \citenamefont {Dzardanov}, \citenamefont {Williams},\ and\
  \citenamefont {Sobolewski}}]{Goltsman2001a}%
  \BibitemOpen
  \bibfield  {author} {\bibinfo {author} {\bibfnamefont {G.~N.}\ \bibnamefont
  {Gol'tsman}}, \bibinfo {author} {\bibfnamefont {O.}~\bibnamefont {Okunev}},
  \bibinfo {author} {\bibfnamefont {G.}~\bibnamefont {Chulkova}}, \bibinfo
  {author} {\bibfnamefont {A.}~\bibnamefont {Lipatov}}, \bibinfo {author}
  {\bibfnamefont {A.}~\bibnamefont {Semenov}}, \bibinfo {author} {\bibfnamefont
  {K.}~\bibnamefont {Smirnov}}, \bibinfo {author} {\bibfnamefont
  {B.}~\bibnamefont {Voronov}}, \bibinfo {author} {\bibfnamefont
  {A.}~\bibnamefont {Dzardanov}}, \bibinfo {author} {\bibfnamefont
  {C.}~\bibnamefont {Williams}}, \ and\ \bibinfo {author} {\bibfnamefont
  {R.}~\bibnamefont {Sobolewski}},\ }\href
  {http://scitation.aip.org/content/aip/journal/apl/79/6/10.1063/1.1388868}
  {\bibfield  {journal} {\bibinfo  {journal} {Applied Physics Letters}\
  }\textbf {\bibinfo {volume} {79}},\ \bibinfo {pages} {705} (\bibinfo {year}
  {2001})}\BibitemShut {NoStop}%
\bibitem [{\citenamefont {Hadfield}(2009)}]{Hadfield2009}%
  \BibitemOpen
  \bibfield  {author} {\bibinfo {author} {\bibfnamefont {R.~H.}\ \bibnamefont
  {Hadfield}},\ }\href {http://dx.doi.org/10.1038/nphoton.2009.230} {\bibfield
  {journal} {\bibinfo  {journal} {Nat Photon}\ }\textbf {\bibinfo {volume}
  {3}},\ \bibinfo {pages} {696} (\bibinfo {year} {2009})}\BibitemShut {NoStop}%
\bibitem [{\citenamefont {Takesue}\ \emph {et~al.}(2007)\citenamefont
  {Takesue}, \citenamefont {Nam}, \citenamefont {Zhang}, \citenamefont
  {Hadfield}, \citenamefont {Honjo}, \citenamefont {Tamaki},\ and\
  \citenamefont {Yamamoto}}]{Takesue2007}%
  \BibitemOpen
  \bibfield  {author} {\bibinfo {author} {\bibfnamefont {H.}~\bibnamefont
  {Takesue}}, \bibinfo {author} {\bibfnamefont {S.~W.}\ \bibnamefont {Nam}},
  \bibinfo {author} {\bibfnamefont {Q.}~\bibnamefont {Zhang}}, \bibinfo
  {author} {\bibfnamefont {R.~H.}\ \bibnamefont {Hadfield}}, \bibinfo {author}
  {\bibfnamefont {T.}~\bibnamefont {Honjo}}, \bibinfo {author} {\bibfnamefont
  {K.}~\bibnamefont {Tamaki}}, \ and\ \bibinfo {author} {\bibfnamefont
  {Y.}~\bibnamefont {Yamamoto}},\ }\href
  {http://dx.doi.org/10.1038/nphoton.2007.75} {\bibfield  {journal} {\bibinfo
  {journal} {Nat Photon}\ }\textbf {\bibinfo {volume} {1}},\ \bibinfo {pages}
  {343} (\bibinfo {year} {2007})}\BibitemShut {NoStop}%
\bibitem [{\citenamefont {Bussi{\`e}res}\ \emph {et~al.}(2014)\citenamefont
  {Bussi{\`e}res}, \citenamefont {Clausen}, \citenamefont {Tiranov},
  \citenamefont {Korzh}, \citenamefont {Verma}, \citenamefont {Nam},
  \citenamefont {Marsili}, \citenamefont {Ferrier}, \citenamefont {Goldner},
  \citenamefont {Herrmann}, \citenamefont {Silberhorn}, \citenamefont {Sohler},
  \citenamefont {Afzelius},\ and\ \citenamefont {Gisin}}]{Bussieres2014}%
  \BibitemOpen
  \bibfield  {author} {\bibinfo {author} {\bibfnamefont {F.}~\bibnamefont
  {Bussi{\`e}res}}, \bibinfo {author} {\bibfnamefont {C.}~\bibnamefont
  {Clausen}}, \bibinfo {author} {\bibfnamefont {A.}~\bibnamefont {Tiranov}},
  \bibinfo {author} {\bibfnamefont {B.}~\bibnamefont {Korzh}}, \bibinfo
  {author} {\bibfnamefont {V.~B.}\ \bibnamefont {Verma}}, \bibinfo {author}
  {\bibfnamefont {S.~W.}\ \bibnamefont {Nam}}, \bibinfo {author} {\bibfnamefont
  {F.}~\bibnamefont {Marsili}}, \bibinfo {author} {\bibfnamefont
  {A.}~\bibnamefont {Ferrier}}, \bibinfo {author} {\bibfnamefont
  {P.}~\bibnamefont {Goldner}}, \bibinfo {author} {\bibfnamefont
  {H.}~\bibnamefont {Herrmann}}, \bibinfo {author} {\bibfnamefont
  {C.}~\bibnamefont {Silberhorn}}, \bibinfo {author} {\bibfnamefont
  {W.}~\bibnamefont {Sohler}}, \bibinfo {author} {\bibfnamefont
  {M.}~\bibnamefont {Afzelius}}, \ and\ \bibinfo {author} {\bibfnamefont
  {N.}~\bibnamefont {Gisin}},\ }\href
  {http://www.nature.com/nphoton/journal/vaop/ncurrent/full/nphoton.2014.215.html}
  {\bibfield  {journal} {\bibinfo  {journal} {Nat Photon}\ }\textbf {\bibinfo
  {volume} {8}},\ \bibinfo {pages} {775} (\bibinfo {year} {2014})}\BibitemShut
  {NoStop}%
\bibitem [{\citenamefont {Shalm}\ \emph {et~al.}(2015)\citenamefont {Shalm},
  \citenamefont {Meyer-Scott}, \citenamefont {Christensen}, \citenamefont
  {Bierhorst}, \citenamefont {Wayne}, \citenamefont {Stevens}, \citenamefont
  {Gerrits}, \citenamefont {Glancy}, \citenamefont {Hamel}, \citenamefont
  {Allman}, \citenamefont {Coakley}, \citenamefont {Dyer}, \citenamefont
  {Hodge}, \citenamefont {Lita}, \citenamefont {Verma}, \citenamefont
  {Lambrocco}, \citenamefont {Tortorici}, \citenamefont {Migdall},
  \citenamefont {Zhang}, \citenamefont {Kumor}, \citenamefont {Farr},
  \citenamefont {Marsili}, \citenamefont {Shaw}, \citenamefont {Stern},
  \citenamefont {Abell\'an}, \citenamefont {Amaya}, \citenamefont {Pruneri},
  \citenamefont {Jennewein}, \citenamefont {Mitchell}, \citenamefont {Kwiat},
  \citenamefont {Bienfang}, \citenamefont {Mirin}, \citenamefont {Knill},\ and\
  \citenamefont {Nam}}]{Shalm2015a}%
  \BibitemOpen
  \bibfield  {author} {\bibinfo {author} {\bibfnamefont {L.~K.}\ \bibnamefont
  {Shalm}}, \bibinfo {author} {\bibfnamefont {E.}~\bibnamefont {Meyer-Scott}},
  \bibinfo {author} {\bibfnamefont {B.~G.}\ \bibnamefont {Christensen}},
  \bibinfo {author} {\bibfnamefont {P.}~\bibnamefont {Bierhorst}}, \bibinfo
  {author} {\bibfnamefont {M.~A.}\ \bibnamefont {Wayne}}, \bibinfo {author}
  {\bibfnamefont {M.~J.}\ \bibnamefont {Stevens}}, \bibinfo {author}
  {\bibfnamefont {T.}~\bibnamefont {Gerrits}}, \bibinfo {author} {\bibfnamefont
  {S.}~\bibnamefont {Glancy}}, \bibinfo {author} {\bibfnamefont {D.~R.}\
  \bibnamefont {Hamel}}, \bibinfo {author} {\bibfnamefont {M.~S.}\ \bibnamefont
  {Allman}}, \bibinfo {author} {\bibfnamefont {K.~J.}\ \bibnamefont {Coakley}},
  \bibinfo {author} {\bibfnamefont {S.~D.}\ \bibnamefont {Dyer}}, \bibinfo
  {author} {\bibfnamefont {C.}~\bibnamefont {Hodge}}, \bibinfo {author}
  {\bibfnamefont {A.~E.}\ \bibnamefont {Lita}}, \bibinfo {author}
  {\bibfnamefont {V.~B.}\ \bibnamefont {Verma}}, \bibinfo {author}
  {\bibfnamefont {C.}~\bibnamefont {Lambrocco}}, \bibinfo {author}
  {\bibfnamefont {E.}~\bibnamefont {Tortorici}}, \bibinfo {author}
  {\bibfnamefont {A.~L.}\ \bibnamefont {Migdall}}, \bibinfo {author}
  {\bibfnamefont {Y.}~\bibnamefont {Zhang}}, \bibinfo {author} {\bibfnamefont
  {D.~R.}\ \bibnamefont {Kumor}}, \bibinfo {author} {\bibfnamefont {W.~H.}\
  \bibnamefont {Farr}}, \bibinfo {author} {\bibfnamefont {F.}~\bibnamefont
  {Marsili}}, \bibinfo {author} {\bibfnamefont {M.~D.}\ \bibnamefont {Shaw}},
  \bibinfo {author} {\bibfnamefont {J.~A.}\ \bibnamefont {Stern}}, \bibinfo
  {author} {\bibfnamefont {C.}~\bibnamefont {Abell\'an}}, \bibinfo {author}
  {\bibfnamefont {W.}~\bibnamefont {Amaya}}, \bibinfo {author} {\bibfnamefont
  {V.}~\bibnamefont {Pruneri}}, \bibinfo {author} {\bibfnamefont
  {T.}~\bibnamefont {Jennewein}}, \bibinfo {author} {\bibfnamefont {M.~W.}\
  \bibnamefont {Mitchell}}, \bibinfo {author} {\bibfnamefont {P.~G.}\
  \bibnamefont {Kwiat}}, \bibinfo {author} {\bibfnamefont {J.~C.}\ \bibnamefont
  {Bienfang}}, \bibinfo {author} {\bibfnamefont {R.~P.}\ \bibnamefont {Mirin}},
  \bibinfo {author} {\bibfnamefont {E.}~\bibnamefont {Knill}}, \ and\ \bibinfo
  {author} {\bibfnamefont {S.~W.}\ \bibnamefont {Nam}},\ }\href
  {http://link.aps.org/doi/10.1103/PhysRevLett.115.250402} {\bibfield
  {journal} {\bibinfo  {journal} {Phys. Rev. Lett.}\ }\textbf {\bibinfo
  {volume} {115}},\ \bibinfo {pages} {250402} (\bibinfo {year}
  {2015})}\BibitemShut {NoStop}%
\bibitem [{\citenamefont {Shaw}\ \emph {et~al.}(2014)\citenamefont {Shaw},
  \citenamefont {Birnbaum}, \citenamefont {Cheng}, \citenamefont {Srinivasan},
  \citenamefont {Quirk}, \citenamefont {Kovalik}, \citenamefont {Biswas},
  \citenamefont {Beyer}, \citenamefont {Marsili}, \citenamefont {Verma},
  \citenamefont {Mirin}, \citenamefont {Nam}, \citenamefont {Stern},\ and\
  \citenamefont {Farr}}]{Shaw14}%
  \BibitemOpen
  \bibfield  {author} {\bibinfo {author} {\bibfnamefont {M.}~\bibnamefont
  {Shaw}}, \bibinfo {author} {\bibfnamefont {K.}~\bibnamefont {Birnbaum}},
  \bibinfo {author} {\bibfnamefont {M.}~\bibnamefont {Cheng}}, \bibinfo
  {author} {\bibfnamefont {M.}~\bibnamefont {Srinivasan}}, \bibinfo {author}
  {\bibfnamefont {K.}~\bibnamefont {Quirk}}, \bibinfo {author} {\bibfnamefont
  {J.}~\bibnamefont {Kovalik}}, \bibinfo {author} {\bibfnamefont
  {A.}~\bibnamefont {Biswas}}, \bibinfo {author} {\bibfnamefont {A.~D.}\
  \bibnamefont {Beyer}}, \bibinfo {author} {\bibfnamefont {F.}~\bibnamefont
  {Marsili}}, \bibinfo {author} {\bibfnamefont {V.}~\bibnamefont {Verma}},
  \bibinfo {author} {\bibfnamefont {R.~P.}\ \bibnamefont {Mirin}}, \bibinfo
  {author} {\bibfnamefont {S.~W.}\ \bibnamefont {Nam}}, \bibinfo {author}
  {\bibfnamefont {J.~A.}\ \bibnamefont {Stern}}, \ and\ \bibinfo {author}
  {\bibfnamefont {W.~H.}\ \bibnamefont {Farr}},\ }in\ \href
  {http://www.osapublishing.org/abstract.cfm?URI=CLEO_SI-2014-SM4J.2} {\emph
  {\bibinfo {booktitle} {CLEO: 2014}}}\ (\bibinfo {year} {2014})\BibitemShut
  {NoStop}%
\bibitem [{\citenamefont {Sprengers}\ \emph {et~al.}(2011)\citenamefont
  {Sprengers}, \citenamefont {Gaggero}, \citenamefont {Sahin}, \citenamefont
  {Jahanmirinejad}, \citenamefont {Frucci}, \citenamefont {Mattioli},
  \citenamefont {Leoni}, \citenamefont {Beetz}, \citenamefont {Lermer},
  \citenamefont {Kamp}, \citenamefont {H{\"o}fling}, \citenamefont {Sanjines},\
  and\ \citenamefont {Fiore}}]{Sprengers2011}%
  \BibitemOpen
  \bibfield  {author} {\bibinfo {author} {\bibfnamefont {J.~P.}\ \bibnamefont
  {Sprengers}}, \bibinfo {author} {\bibfnamefont {A.}~\bibnamefont {Gaggero}},
  \bibinfo {author} {\bibfnamefont {D.}~\bibnamefont {Sahin}}, \bibinfo
  {author} {\bibfnamefont {S.}~\bibnamefont {Jahanmirinejad}}, \bibinfo
  {author} {\bibfnamefont {G.}~\bibnamefont {Frucci}}, \bibinfo {author}
  {\bibfnamefont {F.}~\bibnamefont {Mattioli}}, \bibinfo {author}
  {\bibfnamefont {R.}~\bibnamefont {Leoni}}, \bibinfo {author} {\bibfnamefont
  {J.}~\bibnamefont {Beetz}}, \bibinfo {author} {\bibfnamefont
  {M.}~\bibnamefont {Lermer}}, \bibinfo {author} {\bibfnamefont
  {M.}~\bibnamefont {Kamp}}, \bibinfo {author} {\bibfnamefont {S.}~\bibnamefont
  {H{\"o}fling}}, \bibinfo {author} {\bibfnamefont {R.}~\bibnamefont
  {Sanjines}}, \ and\ \bibinfo {author} {\bibfnamefont {A.}~\bibnamefont
  {Fiore}},\ }\href
  {http://scitation.aip.org/content/aip/journal/apl/99/18/10.1063/1.3657518}
  {\bibfield  {journal} {\bibinfo  {journal} {Applied Physics Letters}\
  }\textbf {\bibinfo {volume} {99}},\ \bibinfo {eid} {181110} (\bibinfo {year}
  {2011})}\BibitemShut {NoStop}%
\bibitem [{\citenamefont {Rath}\ \emph {et~al.}(2015)\citenamefont {Rath},
  \citenamefont {Kahl}, \citenamefont {Ferrari}, \citenamefont {Sproll},
  \citenamefont {Lewes-Malandrakis}, \citenamefont {Brink}, \citenamefont
  {Ilin}, \citenamefont {Siegel}, \citenamefont {Nebel},\ and\ \citenamefont
  {Pernice}}]{Rath2015}%
  \BibitemOpen
  \bibfield  {author} {\bibinfo {author} {\bibfnamefont {P.}~\bibnamefont
  {Rath}}, \bibinfo {author} {\bibfnamefont {O.}~\bibnamefont {Kahl}}, \bibinfo
  {author} {\bibfnamefont {S.}~\bibnamefont {Ferrari}}, \bibinfo {author}
  {\bibfnamefont {F.}~\bibnamefont {Sproll}}, \bibinfo {author} {\bibfnamefont
  {G.}~\bibnamefont {Lewes-Malandrakis}}, \bibinfo {author} {\bibfnamefont
  {D.}~\bibnamefont {Brink}}, \bibinfo {author} {\bibfnamefont
  {K.}~\bibnamefont {Ilin}}, \bibinfo {author} {\bibfnamefont {M.}~\bibnamefont
  {Siegel}}, \bibinfo {author} {\bibfnamefont {C.}~\bibnamefont {Nebel}}, \
  and\ \bibinfo {author} {\bibfnamefont {W.}~\bibnamefont {Pernice}},\ }\href
  {http://dx.doi.org/10.1038/lsa.2015.111} {\bibfield  {journal} {\bibinfo
  {journal} {Light Sci Appl}\ }\textbf {\bibinfo {volume} {4}},\ \bibinfo
  {pages} {e338} (\bibinfo {year} {2015})}\BibitemShut {NoStop}%
\bibitem [{\citenamefont {Marsili}\ \emph {et~al.}(2013)\citenamefont
  {Marsili}, \citenamefont {Verma}, \citenamefont {Stern}, \citenamefont
  {Harrington}, \citenamefont {Lita}, \citenamefont {Gerrits}, \citenamefont
  {Vayshenker}, \citenamefont {Baek}, \citenamefont {Shaw}, \citenamefont
  {Mirin},\ and\ \citenamefont {Nam}}]{marsili2013}%
  \BibitemOpen
  \bibfield  {author} {\bibinfo {author} {\bibfnamefont {F.}~\bibnamefont
  {Marsili}}, \bibinfo {author} {\bibfnamefont {V.~B.}\ \bibnamefont {Verma}},
  \bibinfo {author} {\bibfnamefont {J.~A.}\ \bibnamefont {Stern}}, \bibinfo
  {author} {\bibfnamefont {S.}~\bibnamefont {Harrington}}, \bibinfo {author}
  {\bibfnamefont {A.~E.}\ \bibnamefont {Lita}}, \bibinfo {author}
  {\bibfnamefont {T.}~\bibnamefont {Gerrits}}, \bibinfo {author} {\bibfnamefont
  {I.}~\bibnamefont {Vayshenker}}, \bibinfo {author} {\bibfnamefont
  {B.}~\bibnamefont {Baek}}, \bibinfo {author} {\bibfnamefont {M.~D.}\
  \bibnamefont {Shaw}}, \bibinfo {author} {\bibfnamefont {R.~P.}\ \bibnamefont
  {Mirin}}, \ and\ \bibinfo {author} {\bibfnamefont {S.~W.}\ \bibnamefont
  {Nam}},\ }\href {http://dx.doi.org/10.1038/nphoton.2013.13} {\bibfield
  {journal} {\bibinfo  {journal} {Nat Photon}\ }\textbf {\bibinfo {volume}
  {7}},\ \bibinfo {pages} {210} (\bibinfo {year} {2013})}\BibitemShut {NoStop}%
\bibitem [{\citenamefont {Korneeva}\ \emph {et~al.}(2014)\citenamefont
  {Korneeva}, \citenamefont {Mikhailov}, \citenamefont {Pershin}, \citenamefont
  {Manova}, \citenamefont {Divochiy}, \citenamefont {Vakhtomin}, \citenamefont
  {Korneev}, \citenamefont {Smirnov}, \citenamefont {Sivakov}, \citenamefont
  {Devizenko},\ and\ \citenamefont {Goltsman}}]{korneeva2014}%
  \BibitemOpen
  \bibfield  {author} {\bibinfo {author} {\bibfnamefont {Y.~P.}\ \bibnamefont
  {Korneeva}}, \bibinfo {author} {\bibfnamefont {M.~Y.}\ \bibnamefont
  {Mikhailov}}, \bibinfo {author} {\bibfnamefont {Y.~P.}\ \bibnamefont
  {Pershin}}, \bibinfo {author} {\bibfnamefont {N.~N.}\ \bibnamefont {Manova}},
  \bibinfo {author} {\bibfnamefont {A.~V.}\ \bibnamefont {Divochiy}}, \bibinfo
  {author} {\bibfnamefont {Y.~B.}\ \bibnamefont {Vakhtomin}}, \bibinfo {author}
  {\bibfnamefont {A.~A.}\ \bibnamefont {Korneev}}, \bibinfo {author}
  {\bibfnamefont {K.~V.}\ \bibnamefont {Smirnov}}, \bibinfo {author}
  {\bibfnamefont {A.~G.}\ \bibnamefont {Sivakov}}, \bibinfo {author}
  {\bibfnamefont {A.~Y.}\ \bibnamefont {Devizenko}}, \ and\ \bibinfo {author}
  {\bibfnamefont {G.~N.}\ \bibnamefont {Goltsman}},\ }\href
  {http://stacks.iop.org/0953-2048/27/i=9/a=095012} {\bibfield  {journal}
  {\bibinfo  {journal} {Superconductor Science and Technology}\ }\textbf
  {\bibinfo {volume} {27}},\ \bibinfo {pages} {095012} (\bibinfo {year}
  {2014})}\BibitemShut {NoStop}%
\bibitem [{\citenamefont {Verma}\ \emph {et~al.}(2015)\citenamefont {Verma},
  \citenamefont {Korzh}, \citenamefont {Bussi\`{e}res}, \citenamefont
  {Horansky}, \citenamefont {Dyer}, \citenamefont {Lita}, \citenamefont
  {Vayshenker}, \citenamefont {Marsili}, \citenamefont {Shaw}, \citenamefont
  {Zbinden}, \citenamefont {Mirin},\ and\ \citenamefont {Nam}}]{Verma2015}%
  \BibitemOpen
  \bibfield  {author} {\bibinfo {author} {\bibfnamefont {V.~B.}\ \bibnamefont
  {Verma}}, \bibinfo {author} {\bibfnamefont {B.}~\bibnamefont {Korzh}},
  \bibinfo {author} {\bibfnamefont {F.}~\bibnamefont {Bussi\`{e}res}}, \bibinfo
  {author} {\bibfnamefont {R.~D.}\ \bibnamefont {Horansky}}, \bibinfo {author}
  {\bibfnamefont {S.~D.}\ \bibnamefont {Dyer}}, \bibinfo {author}
  {\bibfnamefont {A.~E.}\ \bibnamefont {Lita}}, \bibinfo {author}
  {\bibfnamefont {I.}~\bibnamefont {Vayshenker}}, \bibinfo {author}
  {\bibfnamefont {F.}~\bibnamefont {Marsili}}, \bibinfo {author} {\bibfnamefont
  {M.~D.}\ \bibnamefont {Shaw}}, \bibinfo {author} {\bibfnamefont
  {H.}~\bibnamefont {Zbinden}}, \bibinfo {author} {\bibfnamefont {R.~P.}\
  \bibnamefont {Mirin}}, \ and\ \bibinfo {author} {\bibfnamefont {S.~W.}\
  \bibnamefont {Nam}},\ }\href
  {http://www.opticsexpress.org/abstract.cfm?URI=oe-23-26-33792} {\bibfield
  {journal} {\bibinfo  {journal} {Opt. Express}\ }\textbf {\bibinfo {volume}
  {23}},\ \bibinfo {pages} {33792} (\bibinfo {year} {2015})}\BibitemShut
  {NoStop}%
\bibitem [{\citenamefont {Verma}\ \emph {et~al.}(2014)\citenamefont {Verma},
  \citenamefont {Lita}, \citenamefont {Vissers}, \citenamefont {Marsili},
  \citenamefont {Pappas}, \citenamefont {Mirin},\ and\ \citenamefont
  {Nam}}]{Verma2014b}%
  \BibitemOpen
  \bibfield  {author} {\bibinfo {author} {\bibfnamefont {V.~B.}\ \bibnamefont
  {Verma}}, \bibinfo {author} {\bibfnamefont {A.~E.}\ \bibnamefont {Lita}},
  \bibinfo {author} {\bibfnamefont {M.~R.}\ \bibnamefont {Vissers}}, \bibinfo
  {author} {\bibfnamefont {F.}~\bibnamefont {Marsili}}, \bibinfo {author}
  {\bibfnamefont {D.~P.}\ \bibnamefont {Pappas}}, \bibinfo {author}
  {\bibfnamefont {R.~P.}\ \bibnamefont {Mirin}}, \ and\ \bibinfo {author}
  {\bibfnamefont {S.~W.}\ \bibnamefont {Nam}},\ }\href
  {http://scitation.aip.org/content/aip/journal/apl/105/2/10.1063/1.4890277}
  {\bibfield  {journal} {\bibinfo  {journal} {Applied Physics Letters}\
  }\textbf {\bibinfo {volume} {105}},\ \bibinfo {eid} {022602} (\bibinfo {year}
  {2014})}\BibitemShut {NoStop}%
\bibitem [{\citenamefont {Allman}\ \emph {et~al.}(2015)\citenamefont {Allman},
  \citenamefont {Verma}, \citenamefont {Stevens}, \citenamefont {Gerrits},
  \citenamefont {Horansky}, \citenamefont {Lita}, \citenamefont {Marsili},
  \citenamefont {Beyer}, \citenamefont {Shaw}, \citenamefont {Kumor},
  \citenamefont {Mirin},\ and\ \citenamefont {Nam}}]{Allman2015a}%
  \BibitemOpen
  \bibfield  {author} {\bibinfo {author} {\bibfnamefont {M.~S.}\ \bibnamefont
  {Allman}}, \bibinfo {author} {\bibfnamefont {V.~B.}\ \bibnamefont {Verma}},
  \bibinfo {author} {\bibfnamefont {M.}~\bibnamefont {Stevens}}, \bibinfo
  {author} {\bibfnamefont {T.}~\bibnamefont {Gerrits}}, \bibinfo {author}
  {\bibfnamefont {R.~D.}\ \bibnamefont {Horansky}}, \bibinfo {author}
  {\bibfnamefont {A.~E.}\ \bibnamefont {Lita}}, \bibinfo {author}
  {\bibfnamefont {F.}~\bibnamefont {Marsili}}, \bibinfo {author} {\bibfnamefont
  {A.}~\bibnamefont {Beyer}}, \bibinfo {author} {\bibfnamefont {M.~D.}\
  \bibnamefont {Shaw}}, \bibinfo {author} {\bibfnamefont {D.}~\bibnamefont
  {Kumor}}, \bibinfo {author} {\bibfnamefont {R.}~\bibnamefont {Mirin}}, \ and\
  \bibinfo {author} {\bibfnamefont {S.~W.}\ \bibnamefont {Nam}},\ }\href
  {http://scitation.aip.org/content/aip/journal/apl/106/19/10.1063/1.4921318}
  {\bibfield  {journal} {\bibinfo  {journal} {Applied Physics Letters}\
  }\textbf {\bibinfo {volume} {106}},\ \bibinfo {eid} {192601} (\bibinfo {year}
  {2015})}\BibitemShut {NoStop}%
\bibitem [{\citenamefont {Dorenbos}\ \emph {et~al.}(2008)\citenamefont
  {Dorenbos}, \citenamefont {Reiger}, \citenamefont {Perinetti}, \citenamefont
  {Zwiller}, \citenamefont {Zijlstra},\ and\ \citenamefont
  {Klapwijk}}]{Dorenbos2008}%
  \BibitemOpen
  \bibfield  {author} {\bibinfo {author} {\bibfnamefont {S.~N.}\ \bibnamefont
  {Dorenbos}}, \bibinfo {author} {\bibfnamefont {E.~M.}\ \bibnamefont
  {Reiger}}, \bibinfo {author} {\bibfnamefont {U.}~\bibnamefont {Perinetti}},
  \bibinfo {author} {\bibfnamefont {V.}~\bibnamefont {Zwiller}}, \bibinfo
  {author} {\bibfnamefont {T.}~\bibnamefont {Zijlstra}}, \ and\ \bibinfo
  {author} {\bibfnamefont {T.~M.}\ \bibnamefont {Klapwijk}},\ }\href
  {http://scitation.aip.org/content/aip/journal/apl/93/13/10.1063/1.2990646}
  {\bibfield  {journal} {\bibinfo  {journal} {Applied Physics Letters}\
  }\textbf {\bibinfo {volume} {93}},\ \bibinfo {eid} {131101} (\bibinfo {year}
  {2008})}\BibitemShut {NoStop}%
\bibitem [{\citenamefont {Engel}\ \emph {et~al.}(2012)\citenamefont {Engel},
  \citenamefont {Aeschbacher}, \citenamefont {Inderbitzin}, \citenamefont
  {Schilling}, \citenamefont {Il'in}, \citenamefont {Hofherr}, \citenamefont
  {Siegel}, \citenamefont {Semenov},\ and\ \citenamefont
  {H{\"u}bers}}]{Engel2012TaN}%
  \BibitemOpen
  \bibfield  {author} {\bibinfo {author} {\bibfnamefont {A.}~\bibnamefont
  {Engel}}, \bibinfo {author} {\bibfnamefont {A.}~\bibnamefont {Aeschbacher}},
  \bibinfo {author} {\bibfnamefont {K.}~\bibnamefont {Inderbitzin}}, \bibinfo
  {author} {\bibfnamefont {A.}~\bibnamefont {Schilling}}, \bibinfo {author}
  {\bibfnamefont {K.}~\bibnamefont {Il'in}}, \bibinfo {author} {\bibfnamefont
  {M.}~\bibnamefont {Hofherr}}, \bibinfo {author} {\bibfnamefont
  {M.}~\bibnamefont {Siegel}}, \bibinfo {author} {\bibfnamefont
  {A.}~\bibnamefont {Semenov}}, \ and\ \bibinfo {author} {\bibfnamefont
  {H.-W.}\ \bibnamefont {H{\"u}bers}},\ }\href
  {http://scitation.aip.org/content/aip/journal/apl/100/6/10.1063/1.3684243}
  {\bibfield  {journal} {\bibinfo  {journal} {Applied Physics Letters}\
  }\textbf {\bibinfo {volume} {100}},\ \bibinfo {eid} {062601} (\bibinfo {year}
  {2012})}\BibitemShut {NoStop}%
\bibitem [{\citenamefont {Baek}\ \emph {et~al.}(2011)\citenamefont {Baek},
  \citenamefont {Lita}, \citenamefont {Verma},\ and\ \citenamefont
  {Nam}}]{Baek2011a}%
  \BibitemOpen
  \bibfield  {author} {\bibinfo {author} {\bibfnamefont {B.}~\bibnamefont
  {Baek}}, \bibinfo {author} {\bibfnamefont {A.~E.}\ \bibnamefont {Lita}},
  \bibinfo {author} {\bibfnamefont {V.}~\bibnamefont {Verma}}, \ and\ \bibinfo
  {author} {\bibfnamefont {S.~W.}\ \bibnamefont {Nam}},\ }\href
  {http://scitation.aip.org/content/aip/journal/apl/98/25/10.1063/1.3600793}
  {\bibfield  {journal} {\bibinfo  {journal} {Applied Physics Letters}\
  }\textbf {\bibinfo {volume} {98}},\ \bibinfo {eid} {251105} (\bibinfo {year}
  {2011})}\BibitemShut {NoStop}%
\bibitem [{\citenamefont {Renema}\ \emph {et~al.}(2014)\citenamefont {Renema},
  \citenamefont {Gaudio}, \citenamefont {Wang}, \citenamefont {Zhou},
  \citenamefont {Gaggero}, \citenamefont {Mattioli}, \citenamefont {Leoni},
  \citenamefont {Sahin}, \citenamefont {de~Dood}, \citenamefont {Fiore},\ and\
  \citenamefont {van Exter}}]{Renema2014}%
  \BibitemOpen
  \bibfield  {author} {\bibinfo {author} {\bibfnamefont {J.~J.}\ \bibnamefont
  {Renema}}, \bibinfo {author} {\bibfnamefont {R.}~\bibnamefont {Gaudio}},
  \bibinfo {author} {\bibfnamefont {Q.}~\bibnamefont {Wang}}, \bibinfo {author}
  {\bibfnamefont {Z.}~\bibnamefont {Zhou}}, \bibinfo {author} {\bibfnamefont
  {A.}~\bibnamefont {Gaggero}}, \bibinfo {author} {\bibfnamefont
  {F.}~\bibnamefont {Mattioli}}, \bibinfo {author} {\bibfnamefont
  {R.}~\bibnamefont {Leoni}}, \bibinfo {author} {\bibfnamefont
  {D.}~\bibnamefont {Sahin}}, \bibinfo {author} {\bibfnamefont {M.~J.~A.}\
  \bibnamefont {de~Dood}}, \bibinfo {author} {\bibfnamefont {A.}~\bibnamefont
  {Fiore}}, \ and\ \bibinfo {author} {\bibfnamefont {M.~P.}\ \bibnamefont {van
  Exter}},\ }\href {http://link.aps.org/doi/10.1103/PhysRevLett.112.117604}
  {\bibfield  {journal} {\bibinfo  {journal} {Phys. Rev. Lett.}\ }\textbf
  {\bibinfo {volume} {112}},\ \bibinfo {pages} {117604} (\bibinfo {year}
  {2014})}\BibitemShut {NoStop}%
\bibitem [{\citenamefont {Akhlaghi}\ \emph {et~al.}(2011)\citenamefont
  {Akhlaghi}, \citenamefont {Majedi},\ and\ \citenamefont
  {Lundeen}}]{Akhlaghi2011}%
  \BibitemOpen
  \bibfield  {author} {\bibinfo {author} {\bibfnamefont {M.~K.}\ \bibnamefont
  {Akhlaghi}}, \bibinfo {author} {\bibfnamefont {A.~H.}\ \bibnamefont
  {Majedi}}, \ and\ \bibinfo {author} {\bibfnamefont {J.~S.}\ \bibnamefont
  {Lundeen}},\ }\href
  {http://www.opticsexpress.org/abstract.cfm?URI=oe-19-22-21305} {\bibfield
  {journal} {\bibinfo  {journal} {Opt. Express}\ }\textbf {\bibinfo {volume}
  {19}},\ \bibinfo {pages} {21305} (\bibinfo {year} {2011})}\BibitemShut
  {NoStop}%
\bibitem [{\citenamefont {Renema}\ \emph {et~al.}(2015)\citenamefont {Renema},
  \citenamefont {Wang}, \citenamefont {Gaudio}, \citenamefont {Komen},
  \citenamefont {op~'t Hoog}, \citenamefont {Sahin}, \citenamefont {Schilling},
  \citenamefont {van Exter}, \citenamefont {Fiore}, \citenamefont {Engel},\
  and\ \citenamefont {de~Dood}}]{Renema2015}%
  \BibitemOpen
  \bibfield  {author} {\bibinfo {author} {\bibfnamefont {J.~J.}\ \bibnamefont
  {Renema}}, \bibinfo {author} {\bibfnamefont {Q.}~\bibnamefont {Wang}},
  \bibinfo {author} {\bibfnamefont {R.}~\bibnamefont {Gaudio}}, \bibinfo
  {author} {\bibfnamefont {I.}~\bibnamefont {Komen}}, \bibinfo {author}
  {\bibfnamefont {K.}~\bibnamefont {op~'t Hoog}}, \bibinfo {author}
  {\bibfnamefont {D.}~\bibnamefont {Sahin}}, \bibinfo {author} {\bibfnamefont
  {A.}~\bibnamefont {Schilling}}, \bibinfo {author} {\bibfnamefont {M.~P.}\
  \bibnamefont {van Exter}}, \bibinfo {author} {\bibfnamefont {A.}~\bibnamefont
  {Fiore}}, \bibinfo {author} {\bibfnamefont {A.}~\bibnamefont {Engel}}, \ and\
  \bibinfo {author} {\bibfnamefont {M.~J.~A.}\ \bibnamefont {de~Dood}},\
  }\href@noop {} {\bibfield  {journal} {\bibinfo  {journal} {Nano Letters}\
  }\textbf {\bibinfo {volume} {15}},\ \bibinfo {pages} {4541} (\bibinfo {year}
  {2015})}\BibitemShut {NoStop}%
\bibitem [{\citenamefont {Vodolazov}\ \emph {et~al.}(2015)\citenamefont
  {Vodolazov}, \citenamefont {Korneeva}, \citenamefont {Semenov}, \citenamefont
  {Korneev},\ and\ \citenamefont {Goltsman}}]{Vodolazov2015}%
  \BibitemOpen
  \bibfield  {author} {\bibinfo {author} {\bibfnamefont {D.~Y.}\ \bibnamefont
  {Vodolazov}}, \bibinfo {author} {\bibfnamefont {Y.~P.}\ \bibnamefont
  {Korneeva}}, \bibinfo {author} {\bibfnamefont {A.~V.}\ \bibnamefont
  {Semenov}}, \bibinfo {author} {\bibfnamefont {A.~A.}\ \bibnamefont
  {Korneev}}, \ and\ \bibinfo {author} {\bibfnamefont {G.~N.}\ \bibnamefont
  {Goltsman}},\ }\href {http://link.aps.org/doi/10.1103/PhysRevB.92.104503}
  {\bibfield  {journal} {\bibinfo  {journal} {Phys. Rev. B}\ }\textbf {\bibinfo
  {volume} {92}},\ \bibinfo {pages} {104503} (\bibinfo {year}
  {2015})}\BibitemShut {NoStop}%
\bibitem [{\citenamefont {Miller}\ \emph {et~al.}(2011)\citenamefont {Miller},
  \citenamefont {Lita}, \citenamefont {Calkins}, \citenamefont {Vayshenker},
  \citenamefont {Gruber},\ and\ \citenamefont {Nam}}]{Miller2011}%
  \BibitemOpen
  \bibfield  {author} {\bibinfo {author} {\bibfnamefont {A.~J.}\ \bibnamefont
  {Miller}}, \bibinfo {author} {\bibfnamefont {A.~E.}\ \bibnamefont {Lita}},
  \bibinfo {author} {\bibfnamefont {B.}~\bibnamefont {Calkins}}, \bibinfo
  {author} {\bibfnamefont {I.}~\bibnamefont {Vayshenker}}, \bibinfo {author}
  {\bibfnamefont {S.~M.}\ \bibnamefont {Gruber}}, \ and\ \bibinfo {author}
  {\bibfnamefont {S.~W.}\ \bibnamefont {Nam}},\ }\href
  {http://www.opticsexpress.org/abstract.cfm?URI=oe-19-10-9102} {\bibfield
  {journal} {\bibinfo  {journal} {Opt. Express}\ }\textbf {\bibinfo {volume}
  {19}},\ \bibinfo {pages} {9102} (\bibinfo {year} {2011})}\BibitemShut
  {NoStop}%
\bibitem [{\citenamefont {Fano}(1947)}]{Fano1947}%
  \BibitemOpen
  \bibfield  {author} {\bibinfo {author} {\bibfnamefont {U.}~\bibnamefont
  {Fano}},\ }\href {http://link.aps.org/doi/10.1103/PhysRev.72.26} {\bibfield
  {journal} {\bibinfo  {journal} {Phys. Rev.}\ }\textbf {\bibinfo {volume}
  {72}},\ \bibinfo {pages} {26} (\bibinfo {year} {1947})}\BibitemShut {NoStop}%
\bibitem [{\citenamefont {Kozorezov}\ \emph {et~al.}(2008)\citenamefont
  {Kozorezov}, \citenamefont {Wigmore}, \citenamefont {Martin}, \citenamefont
  {Verhoeve},\ and\ \citenamefont {Peacock}}]{Kozorezov2008}%
  \BibitemOpen
  \bibfield  {author} {\bibinfo {author} {\bibfnamefont {A.~G.}\ \bibnamefont
  {Kozorezov}}, \bibinfo {author} {\bibfnamefont {J.~K.}\ \bibnamefont
  {Wigmore}}, \bibinfo {author} {\bibfnamefont {D.}~\bibnamefont {Martin}},
  \bibinfo {author} {\bibfnamefont {P.}~\bibnamefont {Verhoeve}}, \ and\
  \bibinfo {author} {\bibfnamefont {A.}~\bibnamefont {Peacock}},\ }\href
  {http://dx.doi.org/10.1007/s10909-007-9612-6} {\bibfield  {journal} {\bibinfo
   {journal} {Journal of Low Temperature Physics}\ }\textbf {\bibinfo {volume}
  {151}},\ \bibinfo {pages} {51} (\bibinfo {year} {2008})}\BibitemShut
  {NoStop}%
\bibitem [{\citenamefont {Kozorezov}\ \emph
  {et~al.}(2015{\natexlab{a}})\citenamefont {Kozorezov}, \citenamefont
  {Lambert}, \citenamefont {Marsili}, \citenamefont {Stevens}, \citenamefont
  {Verma}, \citenamefont {Stern}, \citenamefont {Horansky}, \citenamefont
  {Dyer}, \citenamefont {Duff}, \citenamefont {Pappas}, \citenamefont {Lita},
  \citenamefont {Shaw}, \citenamefont {Mirin},\ and\ \citenamefont
  {Nam}}]{Kozorezov2015}%
  \BibitemOpen
  \bibfield  {author} {\bibinfo {author} {\bibfnamefont {A.~G.}\ \bibnamefont
  {Kozorezov}}, \bibinfo {author} {\bibfnamefont {C.}~\bibnamefont {Lambert}},
  \bibinfo {author} {\bibfnamefont {F.}~\bibnamefont {Marsili}}, \bibinfo
  {author} {\bibfnamefont {M.~J.}\ \bibnamefont {Stevens}}, \bibinfo {author}
  {\bibfnamefont {V.~B.}\ \bibnamefont {Verma}}, \bibinfo {author}
  {\bibfnamefont {J.~A.}\ \bibnamefont {Stern}}, \bibinfo {author}
  {\bibfnamefont {R.}~\bibnamefont {Horansky}}, \bibinfo {author}
  {\bibfnamefont {S.}~\bibnamefont {Dyer}}, \bibinfo {author} {\bibfnamefont
  {S.}~\bibnamefont {Duff}}, \bibinfo {author} {\bibfnamefont {D.~P.}\
  \bibnamefont {Pappas}}, \bibinfo {author} {\bibfnamefont {A.}~\bibnamefont
  {Lita}}, \bibinfo {author} {\bibfnamefont {M.~D.}\ \bibnamefont {Shaw}},
  \bibinfo {author} {\bibfnamefont {R.~P.}\ \bibnamefont {Mirin}}, \ and\
  \bibinfo {author} {\bibfnamefont {S.~W.}\ \bibnamefont {Nam}},\ }\href
  {http://link.aps.org/doi/10.1103/PhysRevB.92.064504} {\bibfield  {journal}
  {\bibinfo  {journal} {Phys. Rev. B}\ }\textbf {\bibinfo {volume} {92}},\
  \bibinfo {pages} {064504} (\bibinfo {year} {2015}{\natexlab{a}})}\BibitemShut
  {NoStop}%
\bibitem [{\citenamefont {Kozorezov}\ \emph
  {et~al.}(2015{\natexlab{b}})\citenamefont {Kozorezov}, \citenamefont
  {Lambert}, \citenamefont {Marsili}, \citenamefont {Stevens}, \citenamefont
  {Verma}, \citenamefont {Stern}, \citenamefont {Horansky}, \citenamefont
  {Dyer}, \citenamefont {Shaw}, \citenamefont {Mirin},\ and\ \citenamefont
  {Nam}}]{Kozorezov_SPW}%
  \BibitemOpen
  \bibfield  {author} {\bibinfo {author} {\bibfnamefont {A.}~\bibnamefont
  {Kozorezov}}, \bibinfo {author} {\bibfnamefont {C.}~\bibnamefont {Lambert}},
  \bibinfo {author} {\bibfnamefont {F.}~\bibnamefont {Marsili}}, \bibinfo
  {author} {\bibfnamefont {M.}~\bibnamefont {Stevens}}, \bibinfo {author}
  {\bibfnamefont {V.}~\bibnamefont {Verma}}, \bibinfo {author} {\bibfnamefont
  {J.}~\bibnamefont {Stern}}, \bibinfo {author} {\bibfnamefont
  {R.}~\bibnamefont {Horansky}}, \bibinfo {author} {\bibfnamefont
  {S.}~\bibnamefont {Dyer}}, \bibinfo {author} {\bibfnamefont {M.}~\bibnamefont
  {Shaw}}, \bibinfo {author} {\bibfnamefont {R.}~\bibnamefont {Mirin}}, \ and\
  \bibinfo {author} {\bibfnamefont {S.}~\bibnamefont {Nam}},\ }in\ \href@noop
  {} {\emph {\bibinfo {booktitle} {Single-Photon Workshop, Geneva
  Switzerland}}}\ (\bibinfo {year} {2015})\BibitemShut {NoStop}%
\bibitem [{\citenamefont {Vodolazov}(2014)}]{Vodolazov2014}%
  \BibitemOpen
  \bibfield  {author} {\bibinfo {author} {\bibfnamefont {D.~Y.}\ \bibnamefont
  {Vodolazov}},\ }\href {http://link.aps.org/doi/10.1103/PhysRevB.90.054515}
  {\bibfield  {journal} {\bibinfo  {journal} {Phys. Rev. B}\ }\textbf {\bibinfo
  {volume} {90}},\ \bibinfo {pages} {054515} (\bibinfo {year}
  {2014})}\BibitemShut {NoStop}%
\bibitem [{\citenamefont {Renema}(2015)}]{Renema2015b}%
  \BibitemOpen
  \bibfield  {author} {\bibinfo {author} {\bibfnamefont {J.}~\bibnamefont
  {Renema}},\ }\emph {\bibinfo {title} {The physics of nanowire superconducting
  single-photon detectors}},\ \href@noop {} {Ph.D. thesis} (\bibinfo {year}
  {2015})\BibitemShut {NoStop}%
\bibitem [{\citenamefont {Engel}\ and\ \citenamefont
  {Schilling}(2013)}]{engel2013numerical}%
  \BibitemOpen
  \bibfield  {author} {\bibinfo {author} {\bibfnamefont {A.}~\bibnamefont
  {Engel}}\ and\ \bibinfo {author} {\bibfnamefont {A.}~\bibnamefont
  {Schilling}},\ }\href@noop {} {\bibfield  {journal} {\bibinfo  {journal}
  {Journal of Applied Physics}\ }\textbf {\bibinfo {volume} {114}},\ \bibinfo
  {pages} {214501} (\bibinfo {year} {2013})}\BibitemShut {NoStop}%
\bibitem [{\citenamefont {Lusche}\ \emph {et~al.}(2014)\citenamefont {Lusche},
  \citenamefont {Semenov}, \citenamefont {Ilin}, \citenamefont {Siegel},
  \citenamefont {Korneeva}, \citenamefont {Trifonov}, \citenamefont {Korneev},
  \citenamefont {Goltsman}, \citenamefont {Vodolazov},\ and\ \citenamefont
  {H{\"u}bers}}]{Lusche2014}%
  \BibitemOpen
  \bibfield  {author} {\bibinfo {author} {\bibfnamefont {R.}~\bibnamefont
  {Lusche}}, \bibinfo {author} {\bibfnamefont {A.}~\bibnamefont {Semenov}},
  \bibinfo {author} {\bibfnamefont {K.}~\bibnamefont {Ilin}}, \bibinfo {author}
  {\bibfnamefont {M.}~\bibnamefont {Siegel}}, \bibinfo {author} {\bibfnamefont
  {Y.}~\bibnamefont {Korneeva}}, \bibinfo {author} {\bibfnamefont
  {A.}~\bibnamefont {Trifonov}}, \bibinfo {author} {\bibfnamefont
  {A.}~\bibnamefont {Korneev}}, \bibinfo {author} {\bibfnamefont
  {G.}~\bibnamefont {Goltsman}}, \bibinfo {author} {\bibfnamefont
  {D.}~\bibnamefont {Vodolazov}}, \ and\ \bibinfo {author} {\bibfnamefont
  {H.-W.}\ \bibnamefont {H{\"u}bers}},\ }\href
  {http://scitation.aip.org/content/aip/journal/jap/116/4/10.1063/1.4891105}
  {\bibfield  {journal} {\bibinfo  {journal} {Journal of Applied Physics}\
  }\textbf {\bibinfo {volume} {116}},\ \bibinfo {eid} {043906} (\bibinfo {year}
  {2014})}\BibitemShut {NoStop}%
\bibitem [{\citenamefont {Gaudio}\ \emph {et~al.}(2016)\citenamefont {Gaudio},
  \citenamefont {Renema}, \citenamefont {Zhou}, \citenamefont {Verma},
  \citenamefont {Lita}, \citenamefont {Shainline}, \citenamefont {Stevens},
  \citenamefont {Mirin}, \citenamefont {Nam}, \citenamefont {van Exter},
  \citenamefont {de~Dood},\ and\ \citenamefont {Fiore}}]{Gaudio2016}%
  \BibitemOpen
  \bibfield  {author} {\bibinfo {author} {\bibfnamefont {R.}~\bibnamefont
  {Gaudio}}, \bibinfo {author} {\bibfnamefont {J.~J.}\ \bibnamefont {Renema}},
  \bibinfo {author} {\bibfnamefont {Z.}~\bibnamefont {Zhou}}, \bibinfo {author}
  {\bibfnamefont {V.~B.}\ \bibnamefont {Verma}}, \bibinfo {author}
  {\bibfnamefont {A.~E.}\ \bibnamefont {Lita}}, \bibinfo {author}
  {\bibfnamefont {J.}~\bibnamefont {Shainline}}, \bibinfo {author}
  {\bibfnamefont {M.~J.}\ \bibnamefont {Stevens}}, \bibinfo {author}
  {\bibfnamefont {R.~P.}\ \bibnamefont {Mirin}}, \bibinfo {author}
  {\bibfnamefont {S.~W.}\ \bibnamefont {Nam}}, \bibinfo {author} {\bibfnamefont
  {M.~P.}\ \bibnamefont {van Exter}}, \bibinfo {author} {\bibfnamefont
  {M.~J.~A.}\ \bibnamefont {de~Dood}}, \ and\ \bibinfo {author} {\bibfnamefont
  {A.}~\bibnamefont {Fiore}},\ }\href
  {http://scitation.aip.org/content/aip/journal/apl/109/3/10.1063/1.4958687}
  {\bibfield  {journal} {\bibinfo  {journal} {Applied Physics Letters}\
  }\textbf {\bibinfo {volume} {109}},\ \bibinfo {eid} {031101} (\bibinfo {year}
  {2016})}\BibitemShut {NoStop}%
\end{thebibliography}%



\newpage \null \newpage
\appendix
\subsection*{\Large Appendix}

\subsection*{Discriminator setting}
\begin{figure}[h] 
	\includegraphics[width=85mm]{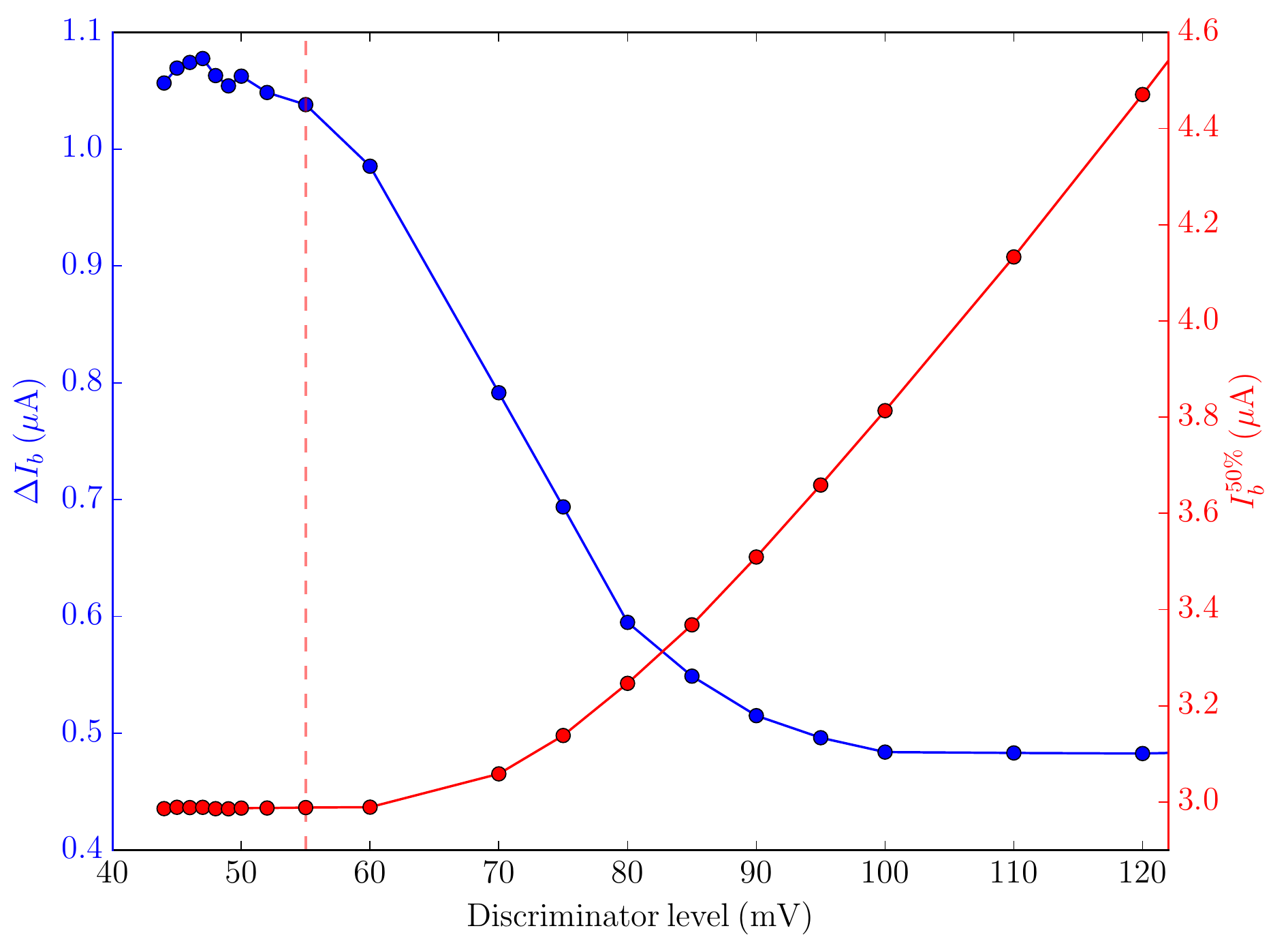}
	\caption{Normalized photon count rate (photon count rate subtracted by the DCR and normalized by the maximum count rate) as a function of the bias current $I_b$ at 0.75~K for an incident photon wavelength of 750~nm. Each color represents one measurement run with a specific discriminator threshold value in millivolts. Each solid line traces the error function fit for the respective data curve. The leftmost and rightmost curves correspond to 44~mV and 150~mV, respectively.}\label{fig:discri_curve_v2}
\end{figure}

\begin{figure}[h] 
	\includegraphics[width=85mm]{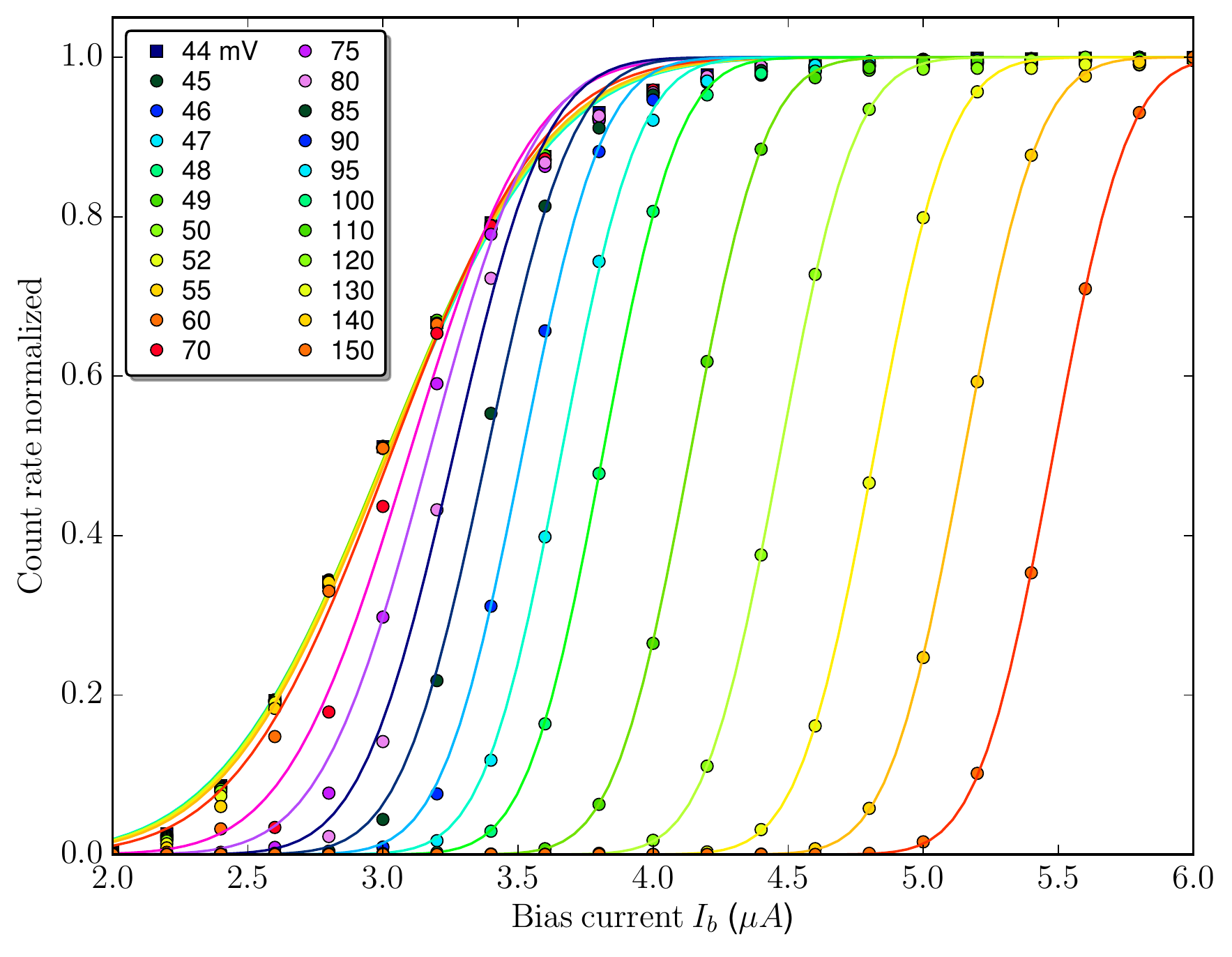}
	\caption{Transition width $\Delta I_b$ (in blue) and $I^{50\%}_{b}$ (in red) in ~$\mathrm{\mu}$A extracted from Fig.~\ref{fig:discri_curve_v2} as a function of the discriminator setting value in mV for an incident photon wavelength of 750~nm. The vertical dashed line represent approximately the maximum setting value of the discriminator.}\label{fig:discri_curve_sigma_mu_v1}
\end{figure}

In order to assess that the experimental data have a physical meaning and are not affected by any electronics effects, we must pay attention to the pulse discrimination electronics. By decreasing the wavelength (increasing the energy) of the incident photon, the bias current needed to create an event decreases. As the signal amplitude depends only on the applied bias current, a problem can arise when the detector is operating at such low currents. If the discriminator threshold level is not set correctly, the consequence is that the photon counts cannot be distinguished from the amplifier noise and the shape of the PCR is affected. In Fig.~\ref{fig:discri_curve_v2}, we plot the PCR as a function of the bias current for different discriminator threshold values at 750~nm. By increasing the threshold value, the bias current needed to overpass it increases and the PCR curves shift to the right. In addition to this, the transition width becomes steeper, see Fig.~\ref{fig:discri_curve_sigma_mu_v1}, where the transition width $\Delta I_b = I_{b}^{80\%} - I_{b}^{20\%}$ and the bias value $I^{50\%}_{b}$ extracted from Fig.~\ref{fig:discri_curve_v2} are plotted as a function of the discriminator level.


We verified that none of the curves are affected by scanning the discriminator level and restricting ourselves to those bias currents where the count rate was independent of threshold level, corresponding the left region shown in Fig.~\ref{fig:discri_curve_sigma_mu_v1}. The minimum detectable voltage pulse in our setup occurs at a bias current of approximately 2.5 $\mu$A.

\subsection*{$\chi^2$ computation}

To check how well our data agree with an error function, we compute the reduced $\chi^2$. We fit our experimental data to the error function of the general form:
\begin{equation}
\mathrm{erf}\left(\frac{I_b-I^{50\%}_b}{\sigma \sqrt{2}}\right) = \frac{2}{\sigma \sqrt{2\pi}} \int_{I^{50\%}_b}^{I_b} \! e^{-\frac{\left(v-I^{50\%}_b\right)^2}{2\sigma^2}} \, \mathrm{d}v
\end{equation}
where $\sigma$ quantifies the width of the PCR transition. 

\begin{figure}[h] 
	\includegraphics[width=85mm]{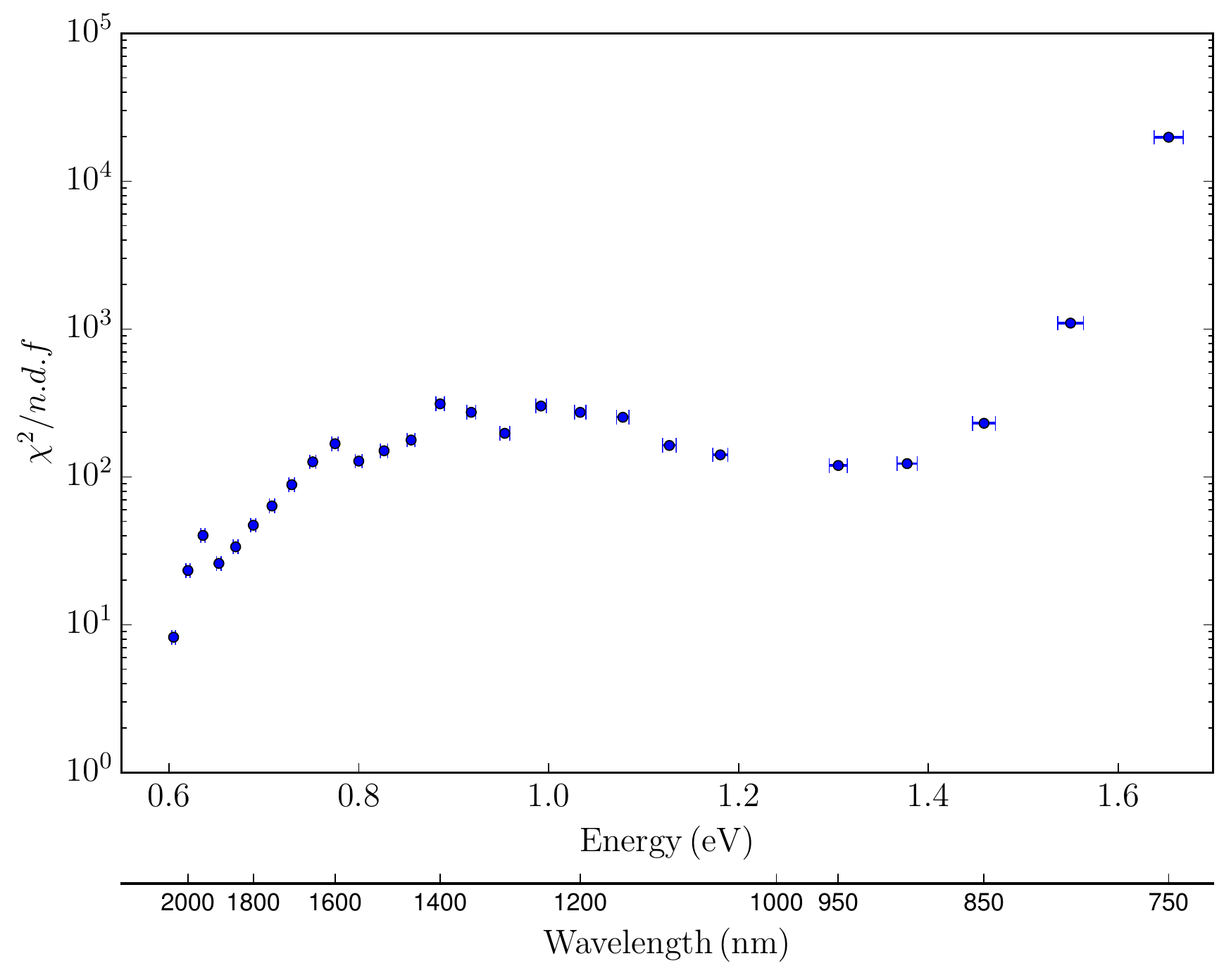}
	\caption{Reduced $\chi^2$ calculated from error function fits on the PCR as a function of the incident photon energy.}\label{fig:energy-chi2_relation}
\end{figure}

We compute the reduced $\chi^2$ given by the following expression:
\begin{equation}
\frac{\chi^2}{\mathrm{n.d.f}} = \sum_{i=0}^{k} \frac{\left(Y_i-\hat{Y}_i \right)^2}{\sigma_i^2} \bigg/ \mathrm{n.d.f}
\end{equation}
where $k$ is the number of points, the $Y_i$ are the observed values with a corresponding standard deviation $\sigma_i$, $\hat{Y_i}$ are the values from the fit, and $\mathrm{n.d.f.}$ is the number of degrees of freedom.

Fig.~\ref{fig:energy-chi2_relation} shows the reduced $\chi^2$ as a function of the incident photon energy. The difference is over two orders of magnitude between the lowest and the highest photon energies, which indicates that the fit at low photon energies agrees much better with the data compared to high energies, for which the shape of the curves starts to deviate from the error function fits.

\end{document}